# M-Net with Bidirectional ConvLSTM for Cup and Disc Segmentation in Fundus Images

Maleeha Khalid Khan
*Department of Software Engineering University of Engineering and Technology* Taxila, Pakistan
maleehakhalidkhan@hotmail.com

Syed Muhammad Anwar
*Department of Software Engineering University of Engineering and Technology* Taxila, Pakistan
s.anwar@uettaxila.edu.pk

*Abstract—* Glaucoma is a severe eye disease that is known to deteriorate optic never fibers, causing cup size to increase, which could result in permanent loss of vision. Glaucoma is the second leading cause of blindness after cataract, but glaucoma being more dangerous as it is not curable. Early diagnoses and treatment of glaucoma can help to slow the progression of glaucoma and its damages. For the detection of glaucoma, the Cup to Disc ratio (CDR) provides significant information. The CDR depends heavily on the accurate segmentation of cup and disc regions. In this paper, we have proposed a modified M-Net with bidirectional convolution long short-term memory (LSTM), based on joint cup and disc segmentation. The proposed network combines features of encoder and decoder, with bidirectional LSTM. Our proposed model segments cup and disc regions based on which the abnormalities in cup to disc ratio can be observed. The proposed model is tested on REFUGE2 data, where our model achieves a dice score of 0.92 for optic disc and an accuracy of 98.99% in segmenting cup and disc regions.

*Keywords— Glaucoma, Deep learning, Segmentation, Glaucoma detection, Cup and disc analysis.*

## I. INTRODUCTION

Glaucoma is a chronic disease which leads to vision loss, and if not treated properly for a prolonged duration the damage caused by glaucoma is not curable. Glaucoma deteriorates the optic nerve fibre, where an early diagnosis and treatment can minimize the damage. For detection of glaucoma, most commonly used assessment methods consider the analysis of optic nerve, measuring intraocular pressure (IOC), and cup to disc ratio (CDR) [1]. In clinical practice, manual methods are used for assessment and detection of glaucoma. Such methods could be time-consuming and require expert knowledge. However, latest techniques such as advancements in machine learning based technologies can help to automate manual assessment methods. In particular, deep learning techniques could help in the detection and clinical diagnosis of glaucoma.

For detection of glaucoma, optic disc (OD) and optic cup (OC) segmentation is commonly used. The cup-to-disc ration (CDR), of optic cup and optic disc provides significant information in detecting glaucoma. While calculating CDR depends heavily on an accurate segmentation of disc to cup regions. Towards this end, we have proposed a modified M-Net [1] architecture, for joint cup and disc segmentation. Our proposed model incorporates bidirectional convolutional long short-term memory (ConvLSTM) module as skip connections. The rest of the paper is organized into the following sections. Section II contained a brief literature review, containing techniques and methods used for detection and analysis of glaucoma. Section III contains the methodology. Section IV contains proposed model, Section V contains experiments and result and Section VI contains the conclusion.

## II. LITERATURE REVIEW

Medical images provide visual information of human body parts which assists in the diagnosis and treatment process. For disease diagnosis using medical images, manual methods have been used to extract useful information. Information from medical images is used in different machine learning techniques, for analysis of medical images and extracting useful information from these images [2]. Machine learning techniques extract important features from medical images and use features to diagnose diseases and make clinical decisions [3].

Recently many deep learning techniques have helped in fast detection of abnormalities, helped in classifying big data, and in segmentation. Deep learning techniques are now used for detecting chronic diseases using data from various imaging modalities such as magnetic resonance imaging (MRI), computed tomography (CT), and ultrasound. Glaucoma is a chronic disease where fundus images are used for the analysis of glaucoma. For detection of glaucoma patient history, eye intra-ocular pressure, disc and cup ratio, and retinal never related information are considered and manual assessments are done. Optical Coherence Tomography (OCT) and Heidelberg Retinal Tomography (HRT) are techniques which produce 2D and 3D visualization of eye. Fundus images provide information about blood vessel, detecting retinal abnormalities, in retinal screening, and optic never details. Deep learning system can use these fundus images for analysis of glaucoma and using these fundus images a system can be trained efficiently to detect glaucoma. The networks work by extracting significant features for detection of glaucoma and these features are later used for classification [4] of glaucoma.

A hybrid color and structure descriptor (HCSD) model [4] was proposed using adaptive histogram equalization and a hybrid combination of feature vectors for early diagnosis of glaucoma. Color fundus images are commonly used in detection of glaucoma, as the area of cup and disc are discriminated by color. A U-Net [5] based fully connected convolution neural network was used to extract features from fundus images. Images were first preprocessed and then segmentation was done using a U-net based model. Moreover, from segmented images false positives were removed using morphological opening. For evaluating the quality of optic disc and optic cup segmentation, self-assessment method use automated diagnosis system. Super pixel-based classification method was used for glaucoma screening and CDR was used for detection of glaucoma, where f-score was shown to improve [6]. The neuro-retinal rim ratio in inferior, superior, temporal and nasal (ISNT) quadrants was considered, where areas of Neuro-retinal Rim (NRR) were used for verification of glaucoma. The hill climbing algorithm was used for the extraction of optic disc, and Fuzzy C-Mean clustering for optic cup extraction from fundus images. The combined effect of these techniques can give better accuracy as compared to manual methods of detection of glaucoma. Glaucoma can also be detected using structural features. Structural changes such as change in retinal layer thickness, changes in optic cup and disc of eye appear earlier as compared to functional features which appear in later stages of glaucoma [7]. Glaucoma can also damage blood vessels in eyes. Optic disc localization and vessel-based segmentation along with SVM classifier helped in detecting retinal abnormalities [8].

Glaucoma causes damage to optic never fiber, which causes the cup size to increase damaging vision. Hence the chances of glaucoma occurrence increase in geriatric population. The main factors that increase the chances of glaucoma are due to increase in intra ocular pressure, genetics, and due to extreme nearsightedness or farsightedness. Earlier detection of glaucoma is necessary, for treatment. The analysis of optic disc, retinal layer, and optic nerve analyses can help in detecting glaucoma. The analysis information helps to detect stage of glaucoma. For treatment of glaucoma ophthalmologist use manual assessment methods which consider assessment of nerve fibers, intraocular pressure and optic cup to disc analysis. The manual assessment methods are not suitable for large population, where automatic detection techniques could play their role using fundus image analysis.

For automatic detection of glaucoma, the central portion is the region of interest that defines the cup area. Cup to Disc ratio analysis is one of the most commonly used method for detection of glaucoma, where a ratio of Cup to disc greater than 0.5 is considered glaucomatous. Deep learning based segmentation methods use separate optic disc and optic cup segmentation methods but these methods ignores the mutual relation of cup and disc, but joint OD and OC joint segmentation [2] helps in preserving features and improve the segmentation. For joint optic disc segmentation model M-Net [3], consisting of a multi-scale input layer and a U-shaped convolution layer and decoder was proposed. M-Net model was based on an end-to-end learning and could detect abnormality in cup size. Visual analysis of eye is done using OCT optical coherence tomography, where OCT images can be used to analyze the tissues in the eye. The abnormal change in tissues texture and change in spatial arrangement of tissues shows glaucoma presence. Region of interest (ROI) is the area focused only on the portion which gives information, which in case of glaucoma can be the blood vessels, cup and disc or retinal never thickness. A deep learning model [4] for detecting blood vessels and thereby classifying the ROI and non-region of interest (NROI) was trained using fundus and gray scale images. The MESSIDOR [9] data was used to classify RIO and NRIO, where ROI helped in defining the area containing useful information for assessment of glaucoma. For biomedical image segmentation, U-Net is a widely used network, an extension of U-Net, Bidirectional ConvLSTM [5] uses dense convolutions instead of simple convolutions to make the network learn more features and performs better segmentation, as compared to a simple LSTM.

## III. METHODOLOGY

Our proposed methodology is an extension of U-Net, where we have used REFUGE2 [6] data for our model training and cross validation. The proposed model has an encoder and decoder, where the encoder side takes input image and performs preprocessing, whereas on the decoder end the image is converted into spatial coordinates. The input fundus images are converted to polar coordinates using polar transformation. This step helps in extracting spatial information from the fundus image. This processed image is then subjected to segmentation using a modified U-net. The model has convolutional layers at encoder side and has bidirectional convolution LSTM between skip connections. Bidirectional LSTM (BLSTM) can learn more spatial features and, and hence it combines features of both the encoding and decoding sides.

In a simple convolutional neural network (CNN) the spatial information is lost when image is segmented. Whereas BLSTM preserves spatial information making segmentation better. Bidirectional LSTM learns more features in both directions and can give better segmentation results as demonstrated by our results. The decoder uses the segmented results and reconverts the images to spatial coordinates. The accuracy of cup and disc segmentation is dependent on the segmentation output.

The step-by-step process performed is as follows:

1. First, the images are pre-processed and transformed into polar coordinates.

2. Second, these polar transformed images are given as input to the deep learning framework.

3. At the encoding side the image size is reduced only to consider the cup and disc portion.

4. The encoding and decoding sides have convolutional layers, pooling layer, ReLU, at the decoding side. Moreover, bidirectional ConvLSTM is added which combines features of encoding and decoding sides.

5. The output from the segmentation model is then converted into spatial coordinates.

## IV. OUR PROPOSED M-BLSTM MODEL

The proposed model (M-BLSTM) is based on modified M-Net, consisting of U-shaped network with Bidirectional convolution LSTM. In the first step, the fundus images are taken as input and pre-processed by converting them into polar coordinates. The processed images are reduced in size focusing only on the region of interest, these images are given as input to the modified U-Net. Images converted into polar coordinates can help to improve segmentation performance by preserving spatial information. The transformed images are given as input to the modified U-Net, for segmentation. U-Net with Bidirectional LSTM between skip connections performs joint cup to disc segmentation.

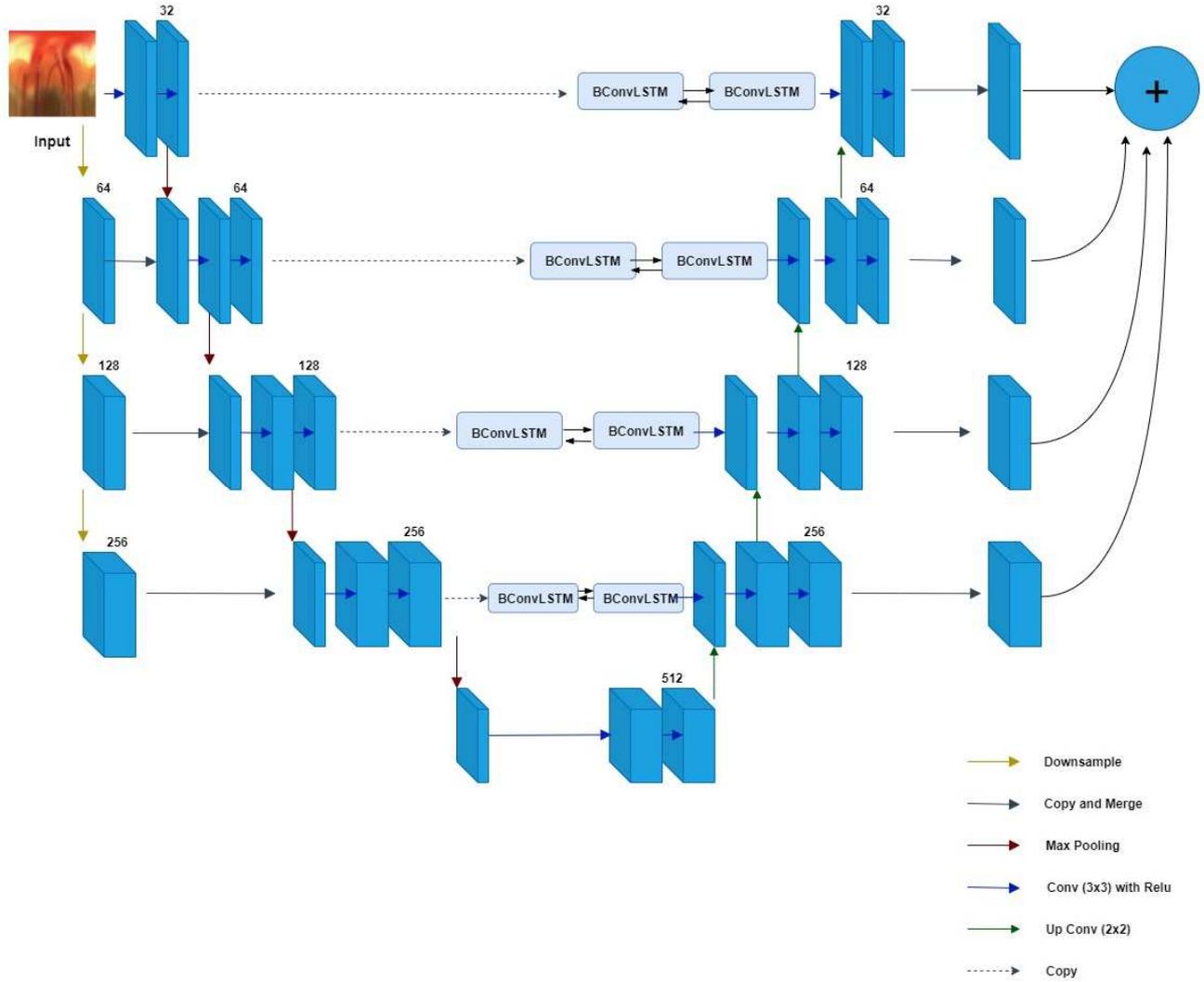

**Fig. 1.** The proposed M-BLSTM Model for segmentation of fundus images, the model takes polar pre-processed images, first, the optic disc is localized, and polar transformed images are generated then M-BLSTM maps for disc and cup region. The model contains convolutional layers along with pooling and up-sampling.

In general LSTM has feedback connections, while simple convolutional LSTM considers one directional information, whereas in bidirectional convolutional LSTM both forward and backward layers information is considered, making network learn more features. On the encoder side the image is transformed into polar coordinates and between convolutional layers ReLU is added, along with pooling and up-sampling. The segmented output is in polar coordinates and on the decoder side images are reconverted into spatial coordinates. From these images the vertical CDR is calculated. Finally, the output from all layers is concatenated and an average segmentation value is calculated. Based on the CDR ratio the stage of glaucoma can be detected, where a higher the value of CDR shows more chances of glaucoma. Figure 1 illustrates our proposed M-BLSTM model.

## V. EXPERIMENTAL SETUP RESULTS

Our proposed model is implemented in python using Keras and with TensorFlow backend. The experiments are performed on Linux 18.04.2 LTS (GNU/Linux). REFUGE2 data are used for disc and cup segmentation along with this the dataset contains test images for glaucoma and non-glaucoma so CDR is calculated and can be cross validated. The REFUGE2 data contains 400 training images, having 40 glaucoma images and 360 non-glaucoma images, which were converted into 2000-disc crop images and 2000 label images in polar coordinates. Disc-crop images were generated using fundus images and label images were generated using ground-truth image. From REFUGE2 data, 400 test images were also used, considering 360 non glaucoma and 40 glaucoma images. From polar transformed images, 80% images were kept for training and 20% for validation. These polar transformed images were subjected to training. After segmentation, based on training results the fundus images were converted into disc-cup mask images. Disc-cup mask images have cup and disc region separated; these resultant images are shown in Figure 3. Disc-cup mask images generated as a result of segmentation and ground-truth images of REFUGE2 data were used to calculate the dice-coefficient. It should be noted that the in REFUGE and REFUGE2 were not balanced. In particular, REFUGE2 data contains only 10% glaucoma images and 90% non-glaucoma images. We believe that due to this high ratio of non-glaucoma images, the segmentation performance could be affected.

Our proposed model shows high accuracy for segmentation. Fundus images from REFUGE2 data are first preprocessed and converted into polar coordinates for each fundus image, polar transformed imaged and ground truth is generated. The polar transformed image gives pixel wise representation of original image in polar coordinates. The polar transformation makes fundus image flat based on optic disc center. In a fundus image in polar coordinate the cup and disc region are layered, and this spatial structure is easy to detect optic disc and cup region. The cup, disc, and background in polar transformed image appears in ordered layer like structure, hence this balanced polar transformed image helps in model training by avoiding overfitting and improves segmentation results. Polar transformation obtains spatial constrains and helps in better optic disc and cup segments. Figure 2 contains glaucoma fundus image, of REFUGE2 data, preprocessed image fundus image in polar coordinates (400×400) image, and ground truth generated in polar coordinates. Also, Figure 2 shows healthy eye fundus image, fundus image in polar coordinates, and ground truth images in polar coordinates. In the ground truth images, the black region represents the background, the orange region represents optic disc, and yellow region represents optic cup. Segmentation results are shown in Figure 3, where the first column shows input fundus images, second column shows the ground truth, the third column shows segmentation using M-Net, and the fourth column show results of segmentation from using out proposed M-BLSTM model.

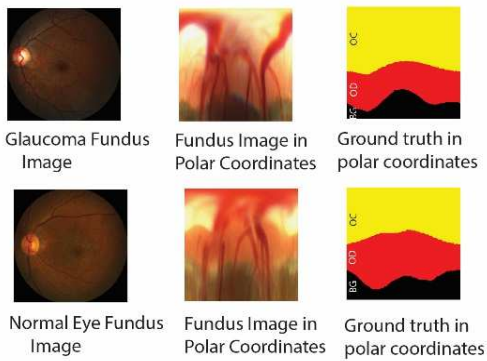

**Fig. 2.** Fundus image in polar coordinates and their ground truth in polar transformed image.

Images after training were subjected to post-processing, where images were converted back to spatial coordinates. In particular, post processing was done by identifying the region of interest, considering cup and disc region. The post processing method requires converting flat polar transformed image into cup disc mask image. The results of segmentation give high accuracy, but post processing of images affects dice coefficient, lowering its value due to the modifications introduced in the cup and disc regions.

The results for experiments performed on REFUGE2 data are presented in Table 1. We have listed average optic cup dice, average optic disc dice, and average validation accuracy of our proposed M-BLSTM model. The average optic cup dice value is calculated by multiplying 2 with overlapping of the segmented image and the ground truth image cup and dividing it with total area. The average dice value is calculated by considering the overlapping area between segmented image and ground truth image and the total area, where the standard dice formula is used. Post processing of images have affected the value of dice coefficient. The average cup and disc dice value shows that our proposed model successfully identifies the disc and cup regions. It should be noted that although, our dice values are lower, this is the first study to the best of our knowledge with such promising results on REFUGE2 data. The increased data size and highly unbalanced class distribution makes the segmentation a challenging problem. In future, the model will be adapted further to perform better on this data. Since the baseline models were designed for another dataset and transfer learning was adopted in this study, making the network hyperparameters generalized for all data is another area of research in retinal image analysis.

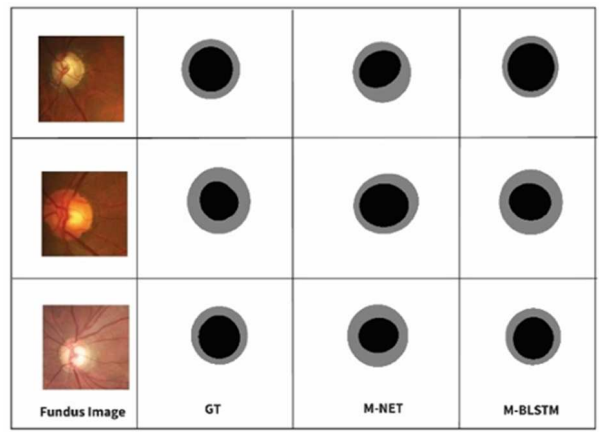

**Fig. 3.** Segmentation results, from left to right fundus image, ground truth, M-Net, and our BLSTM results.

**TABLE I.** PERFORMANCE MEASURE OF OUR PROPOSED M-BLSTM MODEL.

| Sr. # | Method | Data | Optic Cup Avg. Dice | Optic Disc Avg. Dice | Validation Accuracy |
|---|---|---|---|---|---|
| 1 | CUHKMED [10] | REFUGE | 0.88 | 0.96 | ------------ |
| 2 | Masker [10] | REFUGE | 0.88 | 0.94 | ------------ |
| 3 | BUCT [10] | REFUGE | 0.87 | 0.95 | ------------ |
| 4 | NKSG [10] | REFUGE | 0.86 | 0.94 | ------------ |
| 5 | Our proposed M-BLSTM Model | REFUGE2 | 0.86 | 0.92 | 98.99% |

### VI. CONCLUSION

In this paper, we have proposed a modified U-Net based model (M-BLSTM), that is based on joint cup and disc segmentation with bidirectional convolution LSTM. Our proposed model M-BLSTM consist of modified M-Net with bidirectional LSTM added between skip connections, making the network learn more features and improving segmentation results. The proposed M-BLSTM gives segmented disc to cup results, from these segmented results CDR is calculated. The proposed model has been tested on REFUGE2 data where the proposed model achieves a segmentation accuracy of 98.99%.